# Skyrmion Molecule Lattices Enabling Stable Transport and Flexible Manipulation

Lei Liu[1], Xiujuan Zhang[1,*], Ming-Hui Lu[1,2,3,*], and Yan-Feng Chen[1,3,*]

[1]*National Laboratory of Solid State Microstructures and Department of Materials Science and Engineering, Nanjing University, Nanjing 210093, China*

[2]*Jiangsu Key Laboratory of Artificial Functional Materials, Nanjing 210093, China*

[3]*Collaborative Innovation Center of Advanced Microstructures, Nanjing University, Nanjing 210093, China*

*Email: xiujuanzhang@nju.edu.cn, luminghui@nju.edu.cn, yfchen@nju.edu.cn

## Abstract

**Skyrmions—topologically protected nanoscale spin textures with vortex-like configurations—hold transformative potential for ultra-dense data storage, spintronics and quantum computing. However, their practical utility is challenged by dynamic instability, complex interaction, and the lack of deterministic control. While recent efforts using classical wave systems have enabled skyrmion simulations via engineered excitations, these realizations rely on fragile interference patterns, precluding stable transport and flexible control. Here, we introduce a skyrmion molecule lattice, a novel architecture where pairs of spin skyrmions with opposite polarizability are symmetry-locked into stable molecule configurations. These molecules emerge as propagating eigenstates of the system, overcoming the static limitations of previous realizations. We further develop a boundary engineering technique, achieving precise control over skyrmion creation, deformation, annihilation, and polarizability inversion. As a proof of concept, we design a graphene-like acoustic surface wave metamaterial, where meta-atom pairs generate vortices with opposite orbital angular momenta, which couple to acoustic spin textures, forming skyrmion molecules. Experimental measurements confirm their stable transport and flexible control. Our work leverages the symmetry-locked molecule lattice to preserve the topological quasiparticle nature of skyrmions, offering a universal framework for their stabilization, transportation and manipulation. This bridges critical gaps in skyrmion physics, with potential impacts on wave-based sensing, information processing, and topological waveguiding.**

## Introduction

Skyrmions, first conceptualized by Tony Skyrme in 1961 as a particle-like field model describing baryons in nuclear physics[1], were later adapted to condensed matter physics to describe topologically stable swirl or vortex-like spin configurations[2-11]. These structures cannot be smoothly deformed into magnetic ground states with uniformly aligned spins due to their nontrivial topology[8], making them robust against perturbations. This intrinsic stability, combined with their nanoscale size and controllability via external fields such as electric fields[6,7,12-14], magnetic fields[15], optical excitation[16], acoustic waves[17,18], and thermal gradients[19], positions skyrmions as transformative elements for next-generation technologies, from nonvolatile memory to neuromorphic computing. The first experimental observation of magnetic skyrmions was reported in 2009 in the chiral magnet MnSi, where the skyrmions formed a two-dimensional hexagonal lattice[4]. Since then, research into magnetic skyrmions has proliferated, encompassing a wide range of studies from theoretical investigations[8,20-22] to experimental realizations in various materials[23].

Despite progress, key challenges persist. Creating and stabilizing skyrmions requires specific material properties, such as magnetic anisotropies and Dzyaloshinskii-Moriya interactions[24,25], as well as high-purity materials to minimize defects and scattering. These requirements pose significant fabrication challenges. Additionally, the response of skyrmions to the external stimuli is often highly nonlinear, making their dynamic behavior difficult to predict and precise control challenging. On a finer scale, the complex skyrmion interactions, such as the skyrmion Hall effect induced by Magnus force[26,27], lead to trajectory deflection and device-edge annihilation, complicating both fundamental study and device integration. These challenges have motivated a growing interest in clean and controllable classical wave systems as alternative platforms for studying skyrmions[28-41]. Advances in optical[28], acoustic[33] and water wave[41] systems have demonstrated skyrmionic textures and their unique vector dynamics. It has been shown that skyrmion structures are not limited solely to spin configurations, but also manifest in other vector fields, such as electric polarization[28-32,34,39,40] and fluid velocity[37,41], expanding their physical relevance. However, these realizations typically rely on static interference patterns from tailored excitations, which lack the key attributes of intrinsic stability, transportability and flexible control, which are essential for practical applications.

In this work, we overcome these limitations by introducing a skyrmion molecule lattice—a periodic array of bound skyrmion pairs with complementary polarizability, stabilized by lattice symmetry and enabling robust transport and deterministic control. As illustrated in Fig. 1, our approach leverages the interplay between lattice symmetry and anisotropic $p$-orbitals. By strategically arranging the orbitals within a graphene-like lattice, we generate a pair of vortices with opposite chirality at different sublattice sites. These vortices are intrinsically locked to the $K$ and $K'$ valleys and protected by the lattice symmetry, forming a stable vortex molecule. By embedding this vortex molecule lattice in evanescent fields, spin-orbital coupling fuses the out-of-plane spin of the vortices with the in-plane spin of the evanescent fields, generating a full 3D spin vector field and thereby realizing the spin skyrmion molecule lattice. Crucially, these skyrmion molecules emerge as propagating eigenstates of the system, enabling robust transport with minimal scattering. To further address the challenges in skyrmion control, we propose a boundary engineering technique that can modulate skyrmion interactions by fine-tuning sample boundaries. By implementing this technique in an acoustic surface wave metamaterial, we demonstrate the ability to create, deform, annihilate, and even flip the polarizability of skyrmion molecules. Our approach establishes molecule lattices as a versatile platform for stabilizing, transporting and controlling skyrmions, as well as for exploring their interactions in a highly tunable setting.

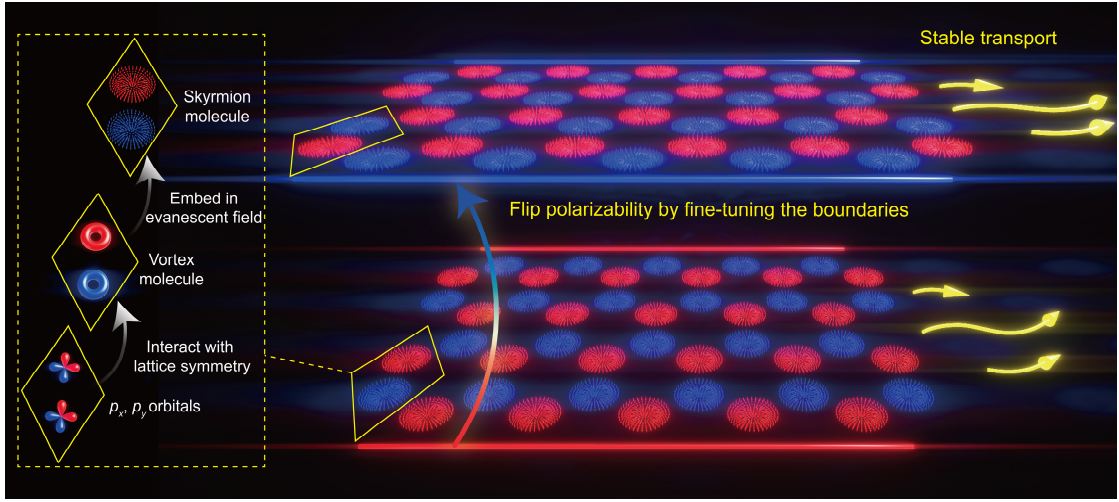

**Fig. 1 | Schematic of the skyrmion molecule lattice.** The left image illustrates the key steps for its realization. Two orthogonal in-pane $p$-orbitals, $|p_x\rangle$ and $|p_y\rangle$, serve as the basis to construct a graphene lattice. Each unit cell (yellow box) contains two sublattice sites. Enabled by graphene's lattice symmetry, $|p_x\rangle$ and $|p_y\rangle$ acquire a 90° phase difference at the Dirac points, superposing into vortices. Due to the inversion symmetry, vortices at different sublattice sites carry opposite topological charges, locked together into a neutral, stable configuration termed a vortex molecule. When coupled to an evanescent field with in-plane spin angular momentum, vortex molecules evolve into skyrmion molecules composed of symmetry-locked skyrmion pairs with opposite skyrmion numbers. As eigenstates of the system, these molecules enable stable transport and flexible control over their creation, deformation, annihilation, and even polarizability inversion by fine-tuning the material boundaries, as depicted in the right image.

## Realization of the skyrmion molecule lattice

As illustrated in Figs. 1 and 2a, we begin with two orthogonal in-plane $p$-orbitals $|p_x\rangle$ and $|p_y\rangle$ as the basis, strategically arranging them to form a graphene-like lattice. This system is governed by transverse ($t_T$) and longitudinal ($t_L$) couplings, where $t_T$ is negligible compared to $t_L$ and set to $t_T = 0$ without loss of generality. In momentum space, the Hamiltonian is given by

$$H(\mathbf{k}) = \begin{bmatrix} \mathbf{0}_{2\times 2} & h(\mathbf{k}) \\ h^\dagger(\mathbf{k}) & \mathbf{0}_{2\times 2} \end{bmatrix}, \tag{1}$$

with

$$h(\mathbf{k}) = t_L \begin{bmatrix} \frac{3}{4}(e^{i\mathbf{k}\cdot\mathbf{e}_2} + e^{i\mathbf{k}\cdot\mathbf{e}_3}) & \frac{\sqrt{3}}{4}(e^{i\mathbf{k}\cdot\mathbf{e}_2} - e^{i\mathbf{k}\cdot\mathbf{e}_3}) \\ \frac{\sqrt{3}}{4}(e^{i\mathbf{k}\cdot\mathbf{e}_2} - e^{i\mathbf{k}\cdot\mathbf{e}_3}) & \frac{1}{4}(e^{i\mathbf{k}\cdot\mathbf{e}_2} + e^{i\mathbf{k}\cdot\mathbf{e}_3}) + e^{i\mathbf{k}\cdot\mathbf{e}_1} \end{bmatrix}. \tag{2}$$

Here, † indicates the complex conjugate transpose. The lattice constant is taken as 1, $\mathbf{k}(k_x, k_y)$ denotes the wave vector, and $\mathbf{e}_n$ ($n = 1, 2, 3$) is the lattice vector in the real space. $H(\mathbf{k})$ corresponds to the eigenfunction $\psi = \left(\phi_{\alpha,p_x}, \phi_{\alpha,p_y}, \phi_{\beta,p_x}, \phi_{\beta,p_y}\right)^T$, where $\phi_{i,p_j}$ ($i = \alpha, \beta$ and $j = x, y$) represents the $|p_j\rangle$ orbital component on site $i$. The strategic orbital arrangement induces distinct propagation phases for $|p_x\rangle$ and $|p_y\rangle$, yielding a total wavefunction $\Psi = |p_x\rangle + e^{i\mathcal{D}}|p_y\rangle$ at each site, where the phase difference $\mathcal{D} = \arg\left(\frac{\phi_{i,p_y}}{\phi_{i,p_x}}\right)$. Taking $t_L = 1$, the band structure incorporating $\mathcal{D}$ is shown in Fig. 2b, which inherits the unique properties of the graphene lattice and features six Dirac points at the $K$ and $K'$ points. At these points, $|p_x\rangle$ and $|p_y\rangle$ acquire a 90° phase difference, with $\mathcal{D} = \frac{\pi}{2}$ for site $\beta$ and $\frac{3\pi}{2}$ for $\alpha$ at the $K$ point, and reversed roles at the $K'$ point due to the time-reversal symmetry. Correspondingly, the orthogonally overlapped $|p_x\rangle$ and $|p_y\rangle$ are superposed, forming circular patterns that manifest as vortices. These vortices exhibit $2\pi$ phase variation around their center, thereby carrying topological charges of $\pm 1$. The corresponding wavefunctions are given by $\Psi_\pm = |p_x\rangle \pm i|p_y\rangle$. The right panel of Fig. 2b illustrates the phase distributions of $|p_x\rangle$ and $|p_y\rangle$ for eigenstates at different Dirac points. At the $K$ point, superposition of $|p_x\rangle$ and $|p_y\rangle$ creates $\Psi_-$ at site $\alpha$ and $\Psi_+$ at $\beta$, which are coupled with each other and locked into even and odd vortex pairs, yielding $M_{K,\text{even}} = \{\Psi_-, \Psi_+\}$ and $M_{K,\text{odd}} = \{\Psi_-, -\Psi_+\}$. At the $K'$ point, the vortex pairs are $M_{K',\text{even}} = \{\Psi_+, \Psi_-\}$ and $M_{K',\text{odd}} = \{\Psi_+, -\Psi_-\}$ (detailed derivations are provided in the supplementary information 1). Notably, these intriguing configurations of vortex pairs are precisely in accordance with the inversion and time-reversal symmetries, reflecting their symmetry-protected nature and intrinsic stability. We designate each stable vortex pair as a molecule.

An in-plane vortex corresponds to a phase singularity in a scalar field, whose spatial gradient induces a rotational vector structure, manifesting as an out-of-plane spin angular momentum[42,43]. To construct a full 3D spin vector field, however, an in-plane spin component is still required. We achieve this by exploiting the evanescent field, whose in-plane and out-of-plane vector components exhibit a $\frac{\pi}{2}$ phase difference. This generates an additional rotation of the field, producing in-plane spin angular momentum[44-46]. By embedding the vortex molecule lattice into such an evanescent environment, the in-plane and out-of-plane spin components interact, resulting in a complete 3D spin vector field. Crucially, the spin vector field inherits the symmetries of the underlying lattice, forming vortex-like, stable spin texture pairs, namely, skyrmion molecules.

To implement the skyrmion molecule lattice, we design an acoustic surface wave metamaterial. As depicted in Fig. 2c, the metamaterial consists of open resonant cavities on a steel plate, forming a graphene-like lattice. Each sublattice site ($\alpha$ or $\beta$) comprises four petaloid cavities hosting two degenerate orthogonal $p$-orbital-like dipole resonant modes, as illustrated by the pressure field distributions in Fig. 2d. These resonant modes form the basis of our design and couple to each other via spoof surface acoustic waves, which are precisely evanescent fields. Figure 2e presents the band structure for this material, featuring Dirac points at the $K$ and $K'$ valleys. To visualize the skyrmion molecules, we show in Fig. 2f (upper panels) the distributions of the acoustic spin angular momentum, $\mathbf{s}$, for the eigenstates at these points. Here, $\mathbf{s}$ describes the rotation of the velocity vector field $\mathbf{v}$, yielding[47]

$$\mathbf{s} = \frac{\rho}{2\omega}\text{Im}(\mathbf{v}^* \times \mathbf{v}), \qquad (3)$$

where $\rho$ denotes the mass density and $\omega$ the angular frequency. It is observed that each molecule indeed contains two spin textures with oppositely swirling configurations, consistent with the typical features of two Néel-type skyrmions $S_\pm$ with skyrmion numbers $\pm 1$ (see Supplementary Information 2 for calculations of the skyrmion numbers). Crucially, the skyrmion molecules strictly align with the symmetry constraints, with one-to-one correspondence to the symmetry-locked vortex molecules, adhering $S_+$ to $\Psi_+$ and $S_-$ to $\Psi_-$. This is confirmed by the lower panels of

Fig. 2f. When examining the scalar features, i.e., the intensity and phase distributions of the pressure field, they directly map to the underlying vortex pairs with opposite chirality. This correlation further highlights the critical role of symmetries in forming and stabilizing skyrmion molecules.

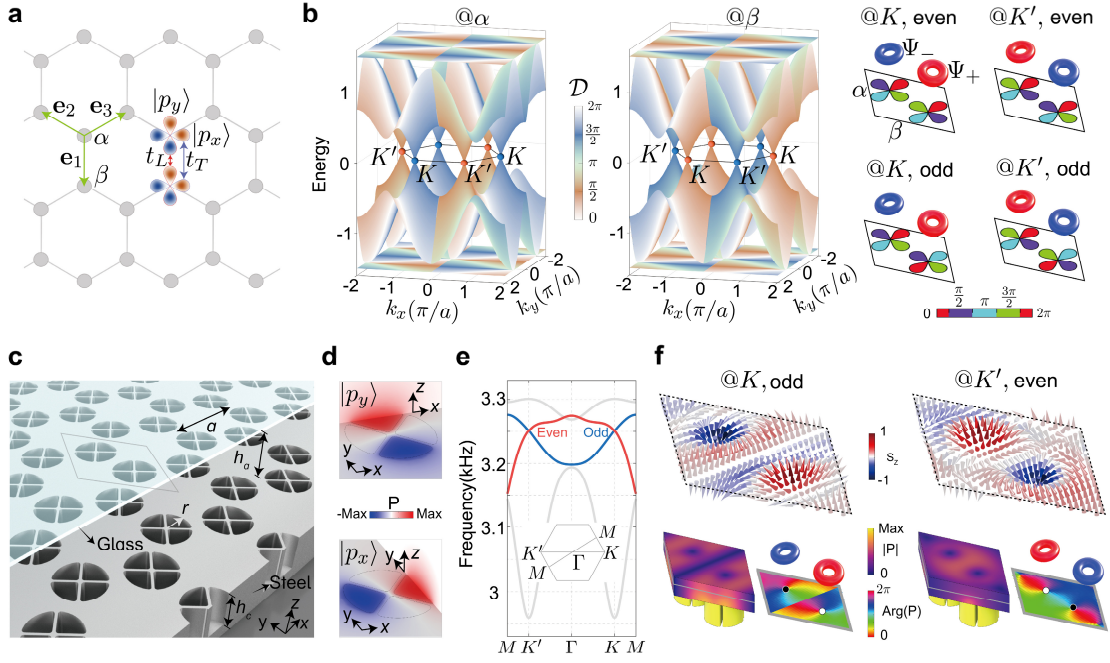

**Fig. 2 | Realization of the skyrmion molecule lattice. a**, Schematic of the $p$-orbital graphene lattice. **b**, Energy bands incorporating $\mathcal{D}$ (the phase difference between $|p_x\rangle$ and $|p_y\rangle$) for sublattice sites $\alpha$ (left) and $\beta$ (middle). At the Dirac points, $90°$ phase difference gives rise to vortices with topological charges of $\pm 1$, yielding wavefunctions $\Psi_\pm = |p_x\rangle \pm i|p_y\rangle$. Enforced and stabilized by the inversion symmetry, these vortices couple with each other and lock into neutral molecules with even and odd parities. Right panel: Eigenstates of the vortex molecules at the $K$ and $K'$ Dirac points. **c**, An acoustic surface wave metamaterial implementing the lattice in **a**. Petaloid cavities are carefully etched on a steel surface to meet the graphene's symmetry requirements while providing an evanescent field. A glass ceiling is placed above the steel surface to confine the surface wave propagation (see Methods for detailed geometric parameters and discussions on the glass ceiling). **d**, Acoustic pressure field distributions of the two dipole resonant modes for each petal, serving as the $|p_x\rangle$ and $|p_y\rangle$ basis. **e**, Band structure of the metamaterial, which hosts Dirac points at the $K$ and $K'$ points, matching the theoretical prediction in **b**. **f**, Normalized spin angular momentum $\hat{\mathbf{s}} = \mathbf{s}/|\mathbf{s}|$ for eigenstates at the Dirac points (upper panels). Emergence of paired vortex-like spin textures with opposite polarizability reveals the skyrmion molecules. Their symmetry correlation with the underlying vortex molecules is confirmed by the pressure field amplitudes and phase distributions (lower panels).

## Observation of robust transport of the skyrmion molecules

Benefiting from our graphene-inspired periodic lattice design, the skyrmion molecules emerge as propagating eigenstates, carrying non-zero group velocity for stable transport. To demonstrate this, we employ boundary engineering, a technique that modulates the material boundaries to select eigenstates satisfying specific boundary conditions, analogous to an infinite potential well[48]. This method, previously used to generate defect-immune bulk states[49,50], is uniquely applied here to isolate and control the even and odd skyrmion molecules. These modes exhibit distinct parities and thus match to different boundary conditions. Figure 3a shows a fabricated waveguide bounded by hard walls along the $y$-direction and open along the $x$-direction (the transport direction). This type of waveguide is commonly used in practice. The hard boundaries enforce zero normal velocity, i.e., $|v_y| = 0$, permitting only the even skyrmion molecule that satisfies this condition, as illustrated by the velocity field distributions in the right panel of Fig. 3a. The odd mode, on the other hand, incompatible with the boundary condition, is excluded. Experimental validation is provided in Fig. 3b by the measured band structure for this waveguide, featuring only the even band (see Methods for the experimental set-up and measurement details).

To observe the transport of the even skyrmion molecule, we launch an excitation from the left port of the waveguide at 3.265 kHz (corresponding to the $K'$ valley). The measured spin vector field (Fig. 3c) reveals stable, uniform $M_{K',\text{even}}$ skyrmion molecules, consistent with both the theory and simulations. Their scalar features are also experimentally confirmed, as shown in Fig. 3d, where paired vortices with chirality $\{+1, -1\}$ are precisely observed. For further evidence, we conduct Fourier analysis on the measured pressure field (see Fig. 3d, right panel), which unambiguously demonstrates $K'$ valley-locking of the molecules—a hallmark of lattice symmetry protection ensuring robust and stable transport (refer to Movies S1 and S2 for time-dependent dynamics and Supplementary Information 3 for more discussions on valley-locking and robustness). Notably, the boundary engineering technique decouples the molecule transport from the waveguide width (along the $y$-direction), enabling arbitrary scaling (see examples in Supplementary Information 4). This scalability, combined with eigenstate-selective control, establishes a platform for flexible and precise skyrmion manipulation.

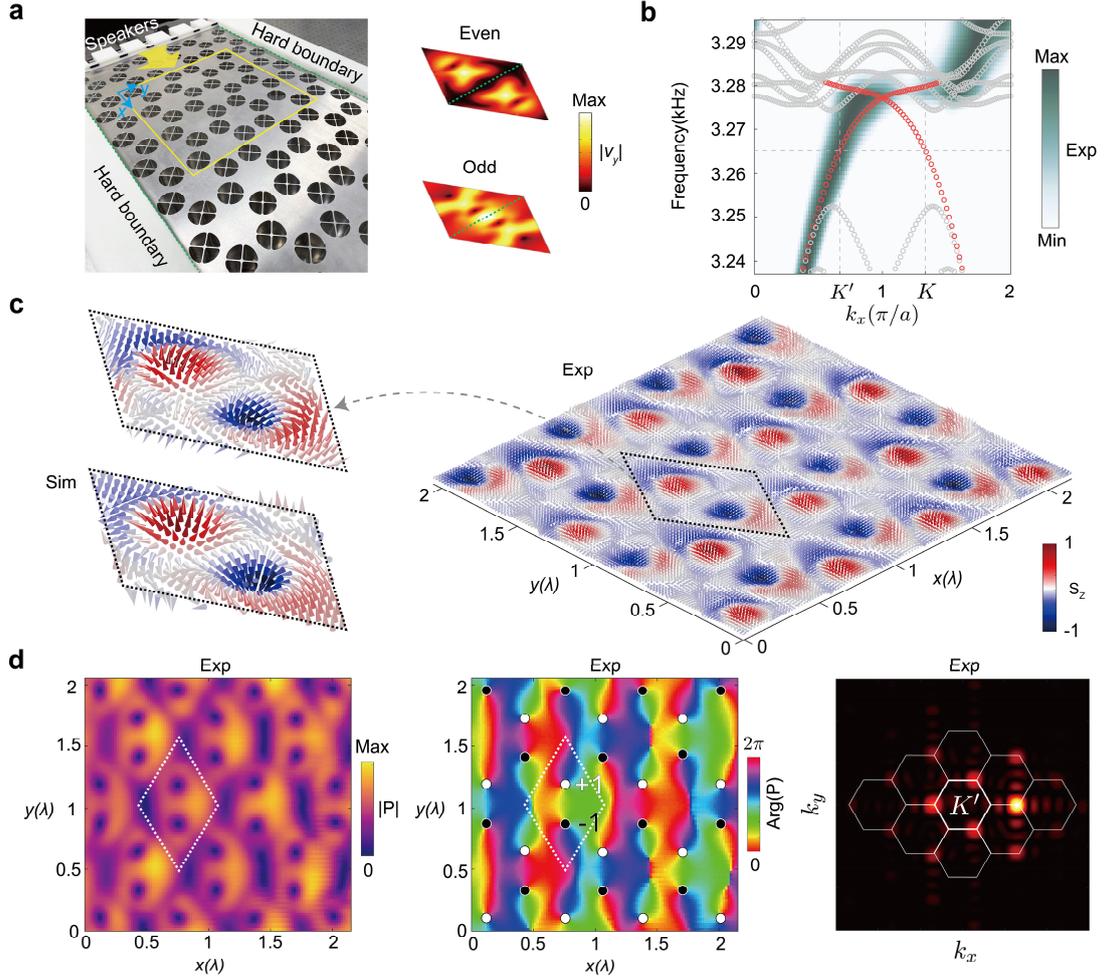

**Fig. 3 | Observation of robust transport of the skyrmion molecules. a**, Left: Fabricated waveguide, open along the *x*-direction and bounded by hard boundaries along the *y*-direction, enforcing zero normal velocity ($|v_y| = 0$). Right: Simulated $|v_y|$ for the even and odd skyrmion molecules. The hard boundary condition admits the even molecule while suppressing the odd one. **b**, Numerically calculated (circles) and experimentally measured (colormap) band structures of the waveguide, validating the boundary selection for the even molecule. **c**, Measured spin vector field $\hat{\mathbf{s}} = \mathbf{s}/|\mathbf{s}|$ (the measurement region is marked by the yellow box in **a**). Regular, uniform field patterns reveal stable skyrmion molecule transport, matching the simulated eigenstate. **d**, Left to right: Measured pressure amplitude $|P|$, phase $\text{Arg}(P)$, and the Fourier spectrum. These results align with both the theoretical and numerical predictions in Fig. 2, highlighting the intrinsic correlation between the skyrmion and vortex molecule lattices and their symmetry-protected valley-locking.

## Deterministic manipulation of the skyrmion molecules

The boundary engineering method enables selection of eigenstates that match specific boundary conditions. By dynamically tuning these conditions, we can manipulate the skyrmion molecules. In our system, the *p*-orbital resonant modes and their coupling via the evanescent field depend on the petaloid cavities. Adjusting their geometry can modify the boundary condition (see Supplementary Information 5 for more details). Experimentally, we deploy the dynamic modulation by injecting measured amounts of silicone oil into the boundary cavities, as illustrated in Fig. 4a. This changes the cavity depths, leading to a smooth and quantitative control over the skyrmion dynamics. Figure

4b tracks the band structure evolution under such boundary modulation, quantified by the cavity depth change $\delta h$. As $\delta h$ increases, the even skyrmion molecules deform, detach from $K$ and $K'$ valley-locking, and annihilate. At $\delta h = 1.2$ mm, the odd skyrmion molecules emerge with reversed polarizability, as the adjusted boundary condition exclusively selects this eigenstate (see Supplementary Information 4 for more details). Such boundary-engineering-driven mode transition is experimentally validated by the measured band structures in Fig. 4b (colormaps).

To corroborate the band structure observation, we further measure the spin vector and scalar pressure fields under the excitation from the left port of the waveguide at 3.265 kHz (see Figs. 4c-d). Again, stable, uniform molecule lattice configurations are observed. Compared to the even mode in Figs. 3c-d, the spin textures here exhibit reversed polarizability and the vortex pairs display inverted phase winding, matching the $M_{K,\text{odd}}$ molecule. Like its even counterpart, the odd molecule is also valley-locked, maintaining stability and robustness during propagation, as confirmed by the Fourier analysis (the right panel of Fig. 4d) and the time-dependent transport dynamics (Movies S3 and S4). Our boundary engineering technique achieves deterministic control over skyrmion molecules, including their creation, deformation, annihilation and polarizability inversion, with high precision and flexible controllability. Integrating this approach with electro-acoustic or electro-optical couplings could potentially unlock real-time, on-chip skyrmion operations in ultrafast and adaptive spin-wave technologies.

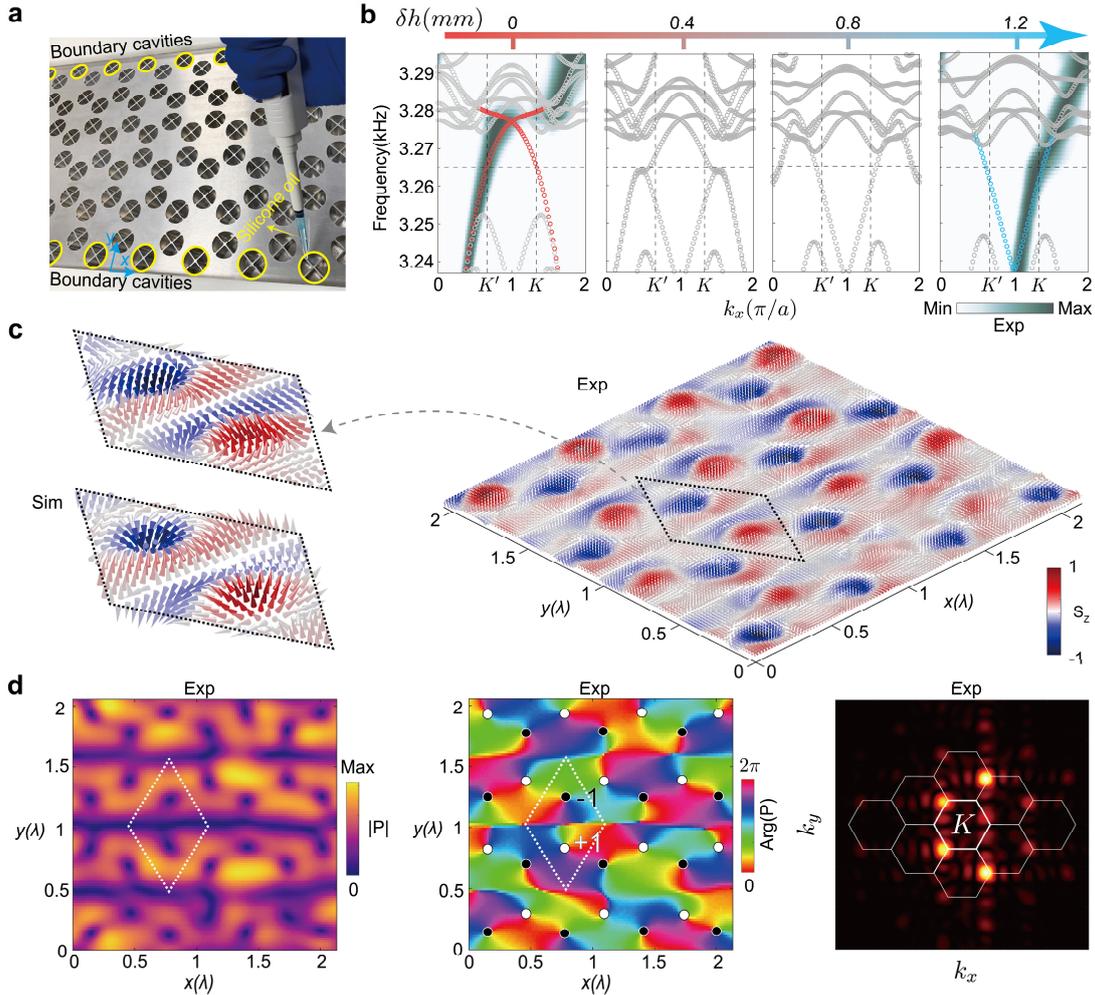

**Fig. 4 | Deterministic skyrmion molecule manipulation via boundary engineering. a**, Boundary modulation by injecting measured amounts of silicone oil into the boundary cavities, tuning their depth. **b**, Band structure evolution under cavity depth change $\delta h$. As $\delta h$ increases, the even band shifts toward high frequencies, detaching from the Dirac points and causing deformation and annihilation of the even skyrmion molecules. Concurrently, the odd band emerges and intersects the Dirac points at $\delta h = 1.2$ mm, signaling the odd skyrmion molecules. **c-d**, Measurements analogous to Figs. 3(c-d), but with boundary cavities modulated by $\delta h = 1.2$ mm. Observation of skyrmion molecules $\{S_-, S_+\}$ and their vortex-like scalar features reveals polarizability inversion relative to the even molecules, validating the boundary engineering control. The Fourier spectrum retains valley-locking, affirming symmetry-protected robustness under boundary engineering.

## Molecule decomposition

The skyrmion molecules comprise symmetry-locked skyrmion pairs with opposite polarizability, resulting in a net topological charge of zero. This neutral, stable configuration preserves the topological quantization of individual skyrmions as quasiparticles. To unveil their quasiparticle nature, we superpose the two ground eigenstates of the system, i.e., the even and odd molecules. This decouples the vortex (and therefore the skyrmion) pairs, isolating their properties at each sublattice, as

$$M_{K,\text{even}} + M_{K,\text{odd}} \to \{\Psi_-, 0\}, \qquad M_{K,\text{even}} - M_{K,\text{odd}} \to \{0, \Psi_+\}, \tag{4a}$$

for the $K$ valley, and

$$M_{K',\text{even}} + M_{K',\text{odd}} \to \{\Psi_+, 0\}, \qquad M_{K',\text{even}} - M_{K',\text{odd}} \to \{0, \Psi_-\}, \tag{4b}$$

for the $K'$ valley.

Figure 5a illustrates the superposition at the $K$ valley using the experimental data in Figs. 3d and 4d. Here, $M_{K,\text{even}}$ is derived by applying the time-reversal operator $\mathcal{T}$ to $M_{K',\text{even}}$, following $M_{K,\text{even}} = \mathcal{T} M_{K',\text{even}}$. The results show that the sublattice $\alpha$ accommodates vortices with topological charge $-1$ while $\beta$ harbors those with $+1$, consistent with Eq. (4a). Correspondingly, the skyrmion molecules decompose into individual skyrmions, as shown in Fig. 5b. We further calculate their skyrmion numbers $n_{sk}$ using

$$n_{sk} = \frac{1}{4\pi} \iint \hat{\mathbf{s}} \cdot \left( \frac{\partial \hat{\mathbf{s}}}{\partial x} \times \frac{\partial \hat{\mathbf{s}}}{\partial y} \right) dx\, dy, \tag{5}$$

where $\hat{\mathbf{s}} = \mathbf{s}/|\mathbf{s}|$ is the normalized spin vector. Experimental data yield $n_{sk} = -0.915$ at $\alpha$ and $n_{sk} = 0.899$ at $\beta$, agreed with the theoretical prediction $\mp 1$ within experimental error (see Methods for error analysis). Superposition at the $K'$ valley is presented in Figs. 5c-d. Due to the time-reversal symmetry, the vortices exhibit exactly reversed chirality compared to the $K$ valley. Concurrently, the skyrmion numbers are inverted, with $n_{sk} = 0.915$ at $\alpha$ and $n_{sk} = -0.899$ at $\beta$. These numbers confirm the topological quantization of the skyrmions, revealing their quasiparticle nature and demonstrating the effectiveness of using molecules to stabilize, transport and manipulate skyrmions.

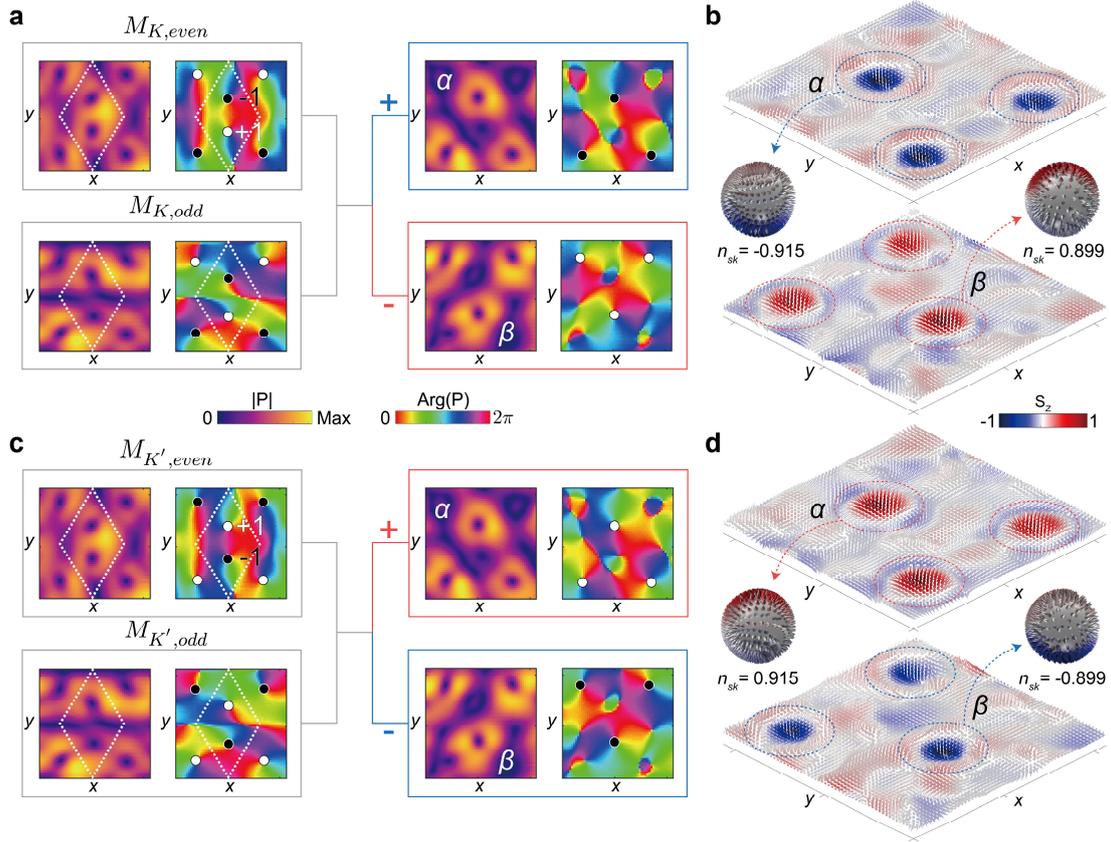

**Fig. 5 | Molecule decomposition into individual skyrmion quasiparticles. a-b**, Superposition of the even and odd molecules at the $K$ valley. For the scalar pressure field superposition in **a**, experimental data in Figs. 3d and 4d are used, while for the spin vector field superposition in **b**, the data in Figs. 3c and 4c are used. The results demonstrate the decoupling of the vortex/skyrmion pairs, with their individual topological properties isolated at different sublattices. This is corroborated by the quantized skyrmion numbers $n_{sk}$ and the stereographic projections of the spin vector field (within the dashed circular regions). **c-d**, The same as **a-b**, but for the $K'$-valley decomposition. Parallel analysis reveals time-reversed configurations.

## Conclusions

We have demonstrated a skyrmion molecule lattice that enables stable transport and precise manipulation, directly addressing critical challenges in skyrmion control. Our method exploits the interactions between anisotropic $p$-orbitals and Bloch momentum in a designed graphene lattice. The lattice symmetry enforces molecule configurations, which effectively stabilize skyrmions while preserving their topological quasiparticle nature. Compared to existing interference methods relying on tailored excitations, the skyrmion molecules in our system emerge as propagating eigenstates with nonzero group velocity, making them inherently compatible with on-chip integration. More crucially, our boundary engineering technique achieves skyrmion creation, deformation, annihilation, and polarizability inversion. Unlike the nonlinear response of magnetic skyrmions to external fields, our skyrmion molecules have linear dependence on the boundary potential which can be accurately controlled by modulating the boundary condition (see Supplementary Information 5). Our method for stabilizing, transporting and controlling skyrmions is universal. In terms of lattice symmetry, it can be generalized to other types of point groups or nonsymmorphic symmetries that enable novel effects like skyrmion Hall effects and non-Abelian skyrmion physics. The $p$-orbital basis can be extended to $d$, $f$, or $g$-orbitals for more sophisticated higher-order spin textures. With the merits of precision and flexibility, the boundary engineering technique can be integrated with active controls such as electro-acoustics, electro-mechanics, or electro-optics, potentially unlocking real-time skyrmion operations in ultrafast and reconfigurable spin-wave technologies. In addition, our method leverages the interaction between vortex lattices and evanescent fields, highlighting the synergy of orbital and spin angular momenta as a scalar-vector duality, which can be important in high-capacity communications, wave-matter interactions, and sensing technologies (see Supplementary Information 6 for an example of using skyrmion molecules to detect the displacement of a small particle, experimentally demonstrating the ability to achieve deep-subwavelength resolution).

## Methods

### Acoustic cavity and lattice design

Supplementary Fig. S9a and Fig. 2c show the details of the designed acoustic open resonant cavity. The cavity has a radius of $r = 1.5$ cm and a depth of $h_c = 2.2$ cm. A block plate divides the cavity into four sections, with a width of $w = 1.5$ mm, while the circular chamfer has a radius of $r_c = 1.5$ mm. A series of resonant modes within this cavity are identified, as shown in Supplementary Fig. S9b. These modes, including those presented in Fig. 2d, are numerically calculated using the 3D acoustic module of the commercial finite-element software COMSOL Multiphysics under eigenfrequency evaluations. The mass density and sound velocity of air are taken as $1.225$ kg·m$^{-3}$ and $341.7$ m·s$^{-1}$, respectively.

The acoustic lattice, shown in Fig. 2c, is designed with a lattice constant of $a = 3.8\sqrt{3}$ cm. A shallower cavity is chosen to achieve a broader bandwidth for the $p$-orbital bands, yet most of them are positioned outside the bound region (i.e., above the sound line) (see Supplementary Information 7). To prevent radiation losses, a narrow propagation channel is constructed between the acrylic ceiling and the steel plate (Fig. 2c), with a channel height of $h_a = 1.6$ cm. This design primarily expands the operational bandwidth while maintaining the evanescent wave field, where intensity decays along the out-of-plane direction.

### Band structure calculations and measurements

For the acoustic lattice in Fig. 2c, we numerically compute its band structure (Fig. 2e) using COMSOL's 3D acoustic module under eigenfrequency evaluations, considering a single unit cell. Additionally, we calculate the projected band along the $x$-direction for the finite lattice in Fig. 3a, which has hard boundaries at both ends along the $y$-direction and periodic boundaries along the $x$-direction. The results are shown in Fig. 3b. When the depths of the boundary cavities are modified (Fig. 4a), the corresponding projected band structures are presented in Fig. 4b.

Experimentally, the sample is fabricated using metal processing and has a length of $12a$. Two 3D-printed photosensitive resin blocks serve the acoustically hard boundary at both ends of the $y$-direction (Fig. 3a). The boundary cavity depths are modified by introducing moderate amounts of silicone oil (Fig. 4a), where a depth variation of $\delta h = 1.2$ mm corresponds to 184 $\mu$L of silicone oil in each petaloid-shaped small cavity. An acoustic loudspeaker array at the sample's left end excites forward-propagating waves to measure the band structures. For achieving high signal-to-noise ratio measurements, we configure the loudspeaker array to generate even and odd sources, which efficiently stimulate the acoustic fields of even and odd molecule lattices, respectively. A homemade acoustic sensor then scans the middle line (along the $x$-direction) of the finite sample in discrete steps of $a$. Data acquisition is performed using a DAQ card (NI 9250 and NI 9234). The collected real-space data are transformed into momentum-space band structures via Fourier transform, with zero-padding employed to enhance resolution. Under different boundary conditions, the measured band structures are presented as color maps in Figs. 3b and 4b.

**Acoustic pressure, velocity and spin field measurements**
Under excitation by the speaker array at 3.265 kHz, we measure the acoustic velocity and pressure fields using a homemade three-dimensional acoustic particle velocity sensor. This sensor scans the designated region of the finite sample (indicated by the yellow box in Fig. 3a) in 0.2 mm increments, recording the velocity components $\mathbf{v}(v_x, v_y, v_z)$ and the acoustic pressure $P$ (for details, see Supplementary Information 8). The measured acoustic velocity and pressure fields are normalized, and the velocity fields are substituted into Eq. (3) to calculate the spin fields $\mathbf{s}$, which is further normalized as $\hat{\mathbf{s}} = \mathbf{s}/|\mathbf{s}|$. The resulting spin fields $\hat{\mathbf{s}}$ are plotted in Figs. 3c for the even mode and Figs. 4c for the odd mode, respectively. The corresponding acoustic pressure distributions are shown in Fig. 3d (even mode) and Fig. 4d (odd mode). The measured acoustic pressure fields are subsequently processed via a two-dimensional Fourier transform to yield the intensity distributions in reciprocal space, illustrated in the right panels of Figs. 3d and 4d. By adding a time-dependent factor $e^{i\omega t}$ into both the acoustic pressure and velocity fields, wave dynamics are visualized in Movies S1 and S2 (even mode) and Movies S3 and S4 (odd mode).

**Superpositions of even and odd molecules**
Vortex and skyrmion molecules can be decomposed into individual vortices and skyrmions through superpositions, enabling observation of their quantized topological properties. Using the measured acoustic pressure data for the even mode $P_{\text{even}}$ (Figs. 3d) and the odd mode $P_{\text{odd}}$ (Fig. 4d), we obtain the superpositions at the $K$ and $K'$ valleys as
$$P_{K,\pm} = P_{\text{even}}^* \pm P_{\text{odd}}, \qquad P_{K',\pm} = P_{\text{even}} \pm P_{\text{odd}}^*. \tag{M1}$$
The resulting superposed acoustic pressure fields $P_{K,\pm}$ and $P_{K',\pm}$ are shown in Figs. 5a and 5c, respectively. Similarly, the velocity fields are obtained by superposing the measured velocity fields of even mode $\mathbf{v}_{\text{even}}$ and odd mode $\mathbf{v}_{\text{odd}}$ as
$$\mathbf{v}_{K,\pm} = \mathbf{v}_{\text{even}}^* \pm \mathbf{v}_{\text{odd}}, \qquad \mathbf{v}_{K',\pm} = \mathbf{v}_{\text{even}} \pm \mathbf{v}_{\text{odd}}^*. \tag{M2}$$
By substituting $\mathbf{v}_{K,\pm}$ and $\mathbf{v}_{K',\pm}$ into Eq. (3), the spin fields are derived, normalized by $\hat{\mathbf{s}} = \mathbf{s}/|\mathbf{s}|$, and shown in Figs. 5b and 5d. Skyrmion numbers are then computed using Eq. (5), with the integration domain denoted by the dashed circles. To improve accuracy, linear interpolation is applied to increase the density of data points. The results align with the ideal skyrmion number of $\pm 1$ (see Supplementary Information 9 for the simulation results), with only minor discrepancies attributed to experimental errors. These errors may arise from several factors. First, due to the viscosity of silicone oil, precisely controlling the amount of silicone oil in each boundary cavity is challenging, leading to deviations in the boundary condition. Consequently, the measured results of the odd molecule lattice exhibit slight discrepancies. Second, inserting the acoustic vector sensor into the waveguide may perturb the local velocity field distribution. Despite these factors, our measurements clearly and convincingly demonstrate the quantized topological properties of individual skyrmions as quasiparticles.

## Data availability

The data reported in the main text and the Supplementary Information are available from the corresponding authors upon reasonable request.

## Acknowledgements

This work are supported by the Key R&D Program of Jiangsu Province (Grant No. BK20232015), the National Natural Science Foundation of China (Grants Nos. 12222407 and 92363001), and the National Key R&D Program of China (Grants Nos. 2023YFA1407700 and 2023YFA1406904).

## Author information

### Authors and Affiliations

**National Laboratory of Solid State Microstructures and Department of Materials Science and Engineering, Nanjing University, Nanjing, China**

Lei Liu, Xiujuan Zhang, Ming-Hui Lu & Yan-Feng Chen

**Jiangsu Key Laboratory of Artificial Functional Materials, Nanjing 210093, China**

Ming-Hui Lu

**Collaborative Innovation Center of Advanced Microstructures, Nanjing University, Nanjing 210093, China**

Ming-Hui Lu & Yan-Feng Chen

### Contributions

L.L. and X.J.Z. conceived the idea. L.L. performed the numerical simulation, the theoretical analyses and the experimental measurements. X.J.Z. and L.L. wrote the manuscript. X.J.Z., M.-H.L. and Y.-F.C. supervised all aspects of this work and managed this project.

### Corresponding authors

Correspondence to Xiujuan Zhang, Ming-Hui Lu or Yan-Feng Chen.

## Ethics declarations

### Competing interests

The authors declare no competing interests.